\pdfoutput=1
\documentclass[reprint,aps,prd]{revtex4-1}
\usepackage{graphicx}
\def\be{\begin{equation}}
\def\ee{\end{equation}}
\def\ba{\begin{array}}
\def\ea{\end{array}}
\def\beqn{\begin{eqnarray}}
\def\eeqn{\end{eqnarray}}

\def\bt{\begin{tabular}}
\def\et{\end{tabular}}
\def\bc{\begin{center}}
\def\ec{\end{center}}

\begin{document}
\title{Estimating matter induced CPT violation in Long-Baseline Neutrino Experiments}
\author{Monika Randhawa}
\email{monika@pu.ac.in}
 \affiliation{University Institute of Engineering and Technology,
Panjab University, Chandigarh, India}
\author{Mandip Singh}
 
\affiliation{Department of
Physics, Centre of Advanced Study, Panjab University, Chandigarh, India.
}
\author{Manmohan Gupta}
\affiliation{Department of
Physics, Centre of Advanced Study, Panjab University, Chandigarh, India.
}
\date{\today}

\begin{abstract}
We examine    matter induced CPT violation effects   in  long baseline electron neutrino appearance experiments in a low energy neutrino factory setup. Assuming CPT invariance in vacuum, the magnitude of CPT violating  asymmetry in matter has been estimated using the exact expressions for the transition probabilities. The dependence of the    asymmetry  on the oscillation parameters like mixing angles, mass squared differences as well as on the Dirac CP violating phase  has been investigated.
\end{abstract}


\maketitle
\section{Introduction}
In particle theory, the discrete symmetries C, P and T have  a central importance. Although   C, P, CP and T are violated \cite{violation},  CPT is a good symmetry \cite{cpt} in the Standard Model, therefore, the fundamental CPT violation may be connected to physics beyond the SM, such as string theory \cite{string,string1}. Experimentally, CPT non-conservation can be probed in the neutrino oscillations, where it would manifest itself by showing different oscillation probabilities for the transitions  $\nu_{\alpha}\rightarrow \nu_{\beta}$ and ${\bar \nu_{\beta}}\rightarrow  {\bar \nu_{\alpha}}$ \cite{cptnu,greenberg}. In this context,  although a 2010 observation of MINOS \cite{minos1} reported tension between $\nu_{\mu}$ and ${\bar\nu_{\mu}}$ oscillation  parameters, suggesting CPT violation,   the difference  was not  observed in their revised results in 2012 \cite{minos2}. Nevertheless, the interest in the search of CPT violation continues \cite{koste}, particularly owing to the increasing precision with which the oscillation parameters are being measured in the current generation of long baseline experiments \cite{{t2k},{nova},{lbne},{lbno}}. 

Even if it is assumed that the CPT invariance  theorem holds good, when neutrinos propagate in a material medium, the matter effects, arising due to interaction of neutrinos with an asymmetric matter, lead to CPT violation in neutrino oscillations, known as extrinsic or fake CPT violation \cite{banul,fakecpt}. The matter effects become all the more important in the long baseline neutrino oscillation experiments, where neutrinos travel a long distance in the earth's matter \cite{{t2k},{nova},{lbne},{lbno}}. These fake effects should be accounted for, while searching for CPT violation.

The matter induced CPT violation has been estimated in some of the papers   in the atmospheric as well as long baseline experiments, primarily by using the approximate analytic expressions for the probabilities for various neutrino oscillation channels  \cite{fakecpt}. The validity of the various approximations depends on the baseline length and the energy of the neutrino, as well as on the mixing angle $\theta_{13}$.   Therefore, keeping in mind the   recently determined   large value of $\theta_{13}$  \cite{theta13}, to which the appearance probabilities are very sensitive, as well as the increased precision in the measurement of other oscillation parameters, it becomes imperative to calculate the probabilities in an exact manner and to update the estimates of CPT asymmetry in neutrino oscillation experiments. This becomes particularly important in view of the large $L$ and $E$ range available to the neutrino in the ongoing and future experiments. 
In this regard, the channel that has been most extensively used to estimate the magnitude of CPT violating parameters is the disappearance channel $\nu_{\mu} \rightarrow \nu_{\mu}$ \cite{banul,fakecpt} as it offers high event rates and  little beam contamination. Further, the neutrino oscillation effects in this channel are large, however, it has been pointed out that the matter effects are rather small in $\nu_{\mu} \rightarrow \nu_{\mu}$ oscillations \cite{banul}. Therefore, to study the effects of matter potential, leading to extrinsic CPT violation,  the sub-dominant channel $\nu_{\mu} \rightarrow \nu_{e}$ looks to be more promising.  Further, this channel is the principle appearance channel available to conventional beams and Superbeams. However, the corresponding CPT conjugate channel  ${\bar \nu_e} \rightarrow {\bar \nu_{\mu}}$ is not going to be explored in the ongoing and forthcoming experiments \cite{{t2k},{nova},{lbne},{lbno}} , as these explore  channels which are CP conjugate of each other.
 In this regard neutrino factories, which are under active consideration \cite{factory} offer a combination of CP and CPT conjugate channels, as both electron as well as muon neutrinos are present in the beam. The challenging task in a neutrino factory is to measure the sign of the charge of the produced
lepton. The sign of a muon charge can
be determined  using a magnetized  iron neutrino detector (MIND) \cite{mind}. The possibility to measure the electron (or positron)
charge with magnetized liquid argon detector has also been explored \cite{lar}.
Neutrino  factories with their high luminosities and low backgrounds
allow to investigate the phenomenon of neutrino oscillations with 
unprecedented accuracy.

Assuming CPT invariance in vacuum,  the purpose of this paper is to investigate the matter induced CPT 
violation effects in the $\nu_{\mu} \rightarrow \nu_{e}$ transitions in four different scenarios
 of  long baseline   neutrino  oscillation experiments: e.g. S1: $L=300$\,Km and $E=1$\,GeV, S2: $L=1300$\,Km and $E=3.5$\,GeV, S3: $L=2300$\,Km and $E=5$\,GeV, S4: $L=3000$\,Km and $E=7$\,GeV, where $L$ is the baseline length and $E$ is the average neutrino energy. The choice of baseline and neutrino energy for the above mentioned scenarios is motivated by the  experiments like T2K \cite{t2k}, LBNE \cite{lbne} and LBNO \cite{lbno} etc.. The energy  is chosen to be below 10\,GeV, as it has been suggested that for the large value of $\theta_{13}$, a low energy neutrino factory (LENF) is better optimized \cite{lenf}.  
The extent of extrinsic  CPT violation in the $\nu_{\mu} \rightarrow \nu_{e}$ transitions  has been studied by calculating the CPT asymmetry  using the exact   neutrino oscillation probability  formulas derived using Cayley-Hamilton formalism   \cite{cayley}. A comparison with the approximate calculations has also been discussed. Further, the dependence of CPT violating    asymmetry on the oscillation parameters like mixing angles, mass squared differences as well as on the Dirac CP violating phase   has been examined.

\section{CPT Violating Asymmetry}
For the flavor transition $\alpha \rightarrow \beta$ ($\alpha, \beta =e,\mu,\tau$),  the CPT violation implies that 
\be   P_{\alpha \beta} \neq
 P_ {{\bar\beta}{\bar\alpha}}\,,
\ee
 where   $P_{\alpha \beta} ( P_ {{\bar\beta}{\bar\alpha}})$ is the probability for the neutrino (antineutrino) flavor transition $\nu_{\alpha} \rightarrow \nu_{\beta}$ (${\bar \nu_{\beta}} \rightarrow {\bar \nu_{\alpha}}$). In the present work, we look for the extrinsic CPT effects in the sub-dominant channel $\nu_{\mu} \rightarrow \nu_e$. The exact expression for the probability  $P_{\alpha\beta}$ is quite lengthy and complicated \cite{cayley,exact}, therefore in the literature, several approximate analytic expressions have been derived \cite{akhm,approx}, wherein the probabilities have been expanded up to first or second order in small parameters like  $\alpha (\equiv \frac{\Delta m_{12}^{2}}{\Delta m_{23}^{2}},\, {\rm the ~ hierarchy~ parameter})$  and/or the reactor mixing angle $\theta_{13}$. In view of the  large value of $\theta_{13}$, expanding the probability only up to first order in $\theta_{13}$, takes the results away from the exact numerical values, particularly in the $L/E$ region relevant for the LBL experiments. Therefore, it is recommended that the probabilities be expanded up to second order in both $\alpha$ as well as sin$\theta_{13} (\equiv s_{13})$. For example, the approximate analytic expression for the probability $P_{\mu e}$ for flavor transitions $\nu_{\mu} \rightarrow \nu_e$ is given as  \cite{akhm},
\begin{eqnarray}
P_{\mu e} &= &\alpha^2  \sin^2 2\theta_{12}  c_{23}^2 \frac{\sin^2
         A\Delta}{A^2} + 4  s_{13}^2  s_{23}^2 \frac{\sin^2
         (A-1)\Delta}{(A-1)^2} \nonumber\\
       &  + & 2  \alpha  s_{13}  \sin 2\theta_{12}  
         \sin2\theta_{23} \cos(\Delta + \delta_{\rm CP}) \times       \nonumber \\
      & &  \frac{\sin
         A\Delta}{A}  \frac{\sin (A-1)\Delta}{A-1} \,.
\label{pmue}\end{eqnarray}\\
Similarly, the probability $P_ {{\bar e}{\bar\mu}}$ for the CPT conjugate flavor  transition (${\bar \nu_e} \rightarrow {\bar \nu_{\mu}}$) is given as
\begin{eqnarray}
P_ {{\bar e}{\bar\mu}}  &= &\alpha^2  \sin^2 2\theta_{12}  c_{23}^2 \frac{\sin^2
         A\Delta}{A^2} + 4  s_{13}^2  s_{23}^2 \frac{\sin^2
         (A+1)\Delta}{(A+1)^2} \nonumber\\
       &  + & 2  \alpha  s_{13}  \sin 2\theta_{12}  
         \sin2\theta_{23} \cos(\Delta + \delta_{\rm CP}) \times   \nonumber\\ 
 & & \frac{\sin
         A\Delta}{A}  \frac{\sin (A+1)\Delta}{A+1} \,.
\label{pemubar}\end{eqnarray}
In the above expressions (\ref{pmue}) and (\ref{pemubar}), $s_{ij}={\rm sin}\theta_{ij}$, $c_{ij}={\rm cos}\theta_{ij} ~ (ij\equiv 12,23,13)$, $\delta_{CP}$ is the leptonic Dirac CP violation phase and 
\be A =\frac{2EV}{\Delta m_{31}^{2}},~~~
\Delta  = \frac{\Delta m_{31}^{2} L}{4E}, \ee
where $V$ is the matter potential, which gives the charged current contribution of electron neutrinos to the matter potential, $L$ is the baseline length, $E$ is the neutrino energy and $\Delta m_{31}^{2}$ gives the atmospheric mass squared difference.
The CPT invariance implies that in vacuum, the probabilities $P_{\mu e}$ and $P_ {{\bar e}{\bar\mu}}$  are exactly the same, resulting in their difference being zero, i.e.
\be   P_{\mu e} - P_{{\bar e} {\bar\mu}}=0. ~~~~~({\rm in~ vacuum, where}~ A=0) \ee
However in matter, as mentioned earlier, the oscillation probabilities are modified due to interaction of electron neutrinos with matter particles, leading to the fake CPT violation, measured in terms of the CPT asymmetry, given for $\nu_{\mu} \rightarrow \nu_e$ transition as 
\be
A_{\mu e}^{CPT}   = \frac{P_{\mu e} - P_{{\bar e} {\bar\mu}}}{P_{\mu e} + P_{{\bar e} {\bar\mu}}}.
\label{cpt} \ee
Defining the asymmetry as the ratio of probabilities has the advantage that on the level of event rates, the systematic experimental uncertainties cancel out to a large extent.
\section{Inputs}
Before going into the details of the analysis, we would like to
mention some of the essentials pertaining to various inputs. The
inputs for neutrino masses, mixing angles and leptonic Dirac CP violation phase used in the present analysis at  1$\sigma$ C.L.
  are   as below \cite{foglinew},
\be
 \Delta m_{12}^{2} = 7.54^{+0.26}_{-0.22} \times 10^{-5} \rm{eV}^{2},~\Delta m_{23}^{2}=2.43^{+0.06}_{-0.10}\times 10^{-3} \rm{eV}^{2},
 \label{masses} \ee
\be
{\rm sin}^2\,\theta_{12}  =  0.307^{+0.018}_{-0.016} , ~
 {\rm sin}^2\,\theta_{23}  =  0.386^{+0.024}_{-0.021}, \label{angles} \ee 
\be {\rm sin}^2\,\theta_{13} = 0.0241 \pm 0.0025,~ \delta_{CP} =  1.08^{+0.28}_{-0.31} \,\pi.  \label{delta} \ee 

In the present work, we consider the baseline length $L \leq 3000$\,Km, implying that one can assume the neutrinos to be traveling in the constant matter density of the earth's crust.   The matter potential $V$ varies with the density $\rho$ of the matter,  and for earth crust's density ($\rho_{\rm crust} \simeq 3g/cm^3$ ) is given as $V \simeq 11.34 \times 10^{-14}$ eV.

\begin{figure*} 
\caption{(Color online). CPT asymmetry $A_{\mu e}^{CPT}$ plotted as function of neutrino energy $E$, for the baseline lengths corresponding to    four scenarios given in  Table (\ref{tabcpt}). The dotted curves correspond to approximate calculations using equation (\ref{cpt}), whereas the solid curves correspond to the exact numerical calculations. All other input parameters are kept at their best fit values given in equations (\ref{masses}) - (\ref{delta}).}
 \includegraphics {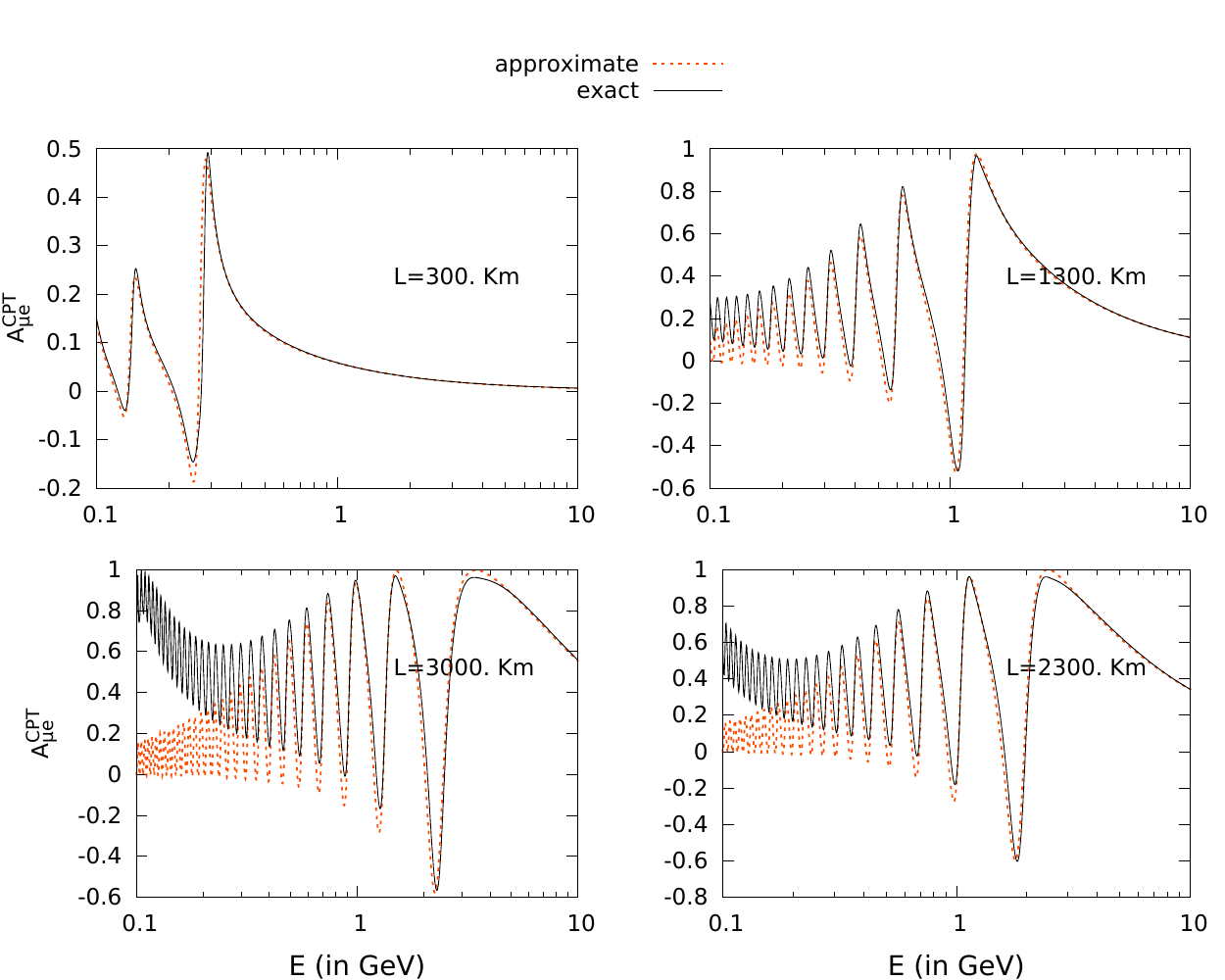}
\label{fige}
\end{figure*}

\section{Numerical Analysis and Results}
Using the exact neutrino oscillation probability  formulas derived using Cayley-Hamilton formalism \cite{cayley} and the  input parameters given in the equations (\ref{masses}) - (\ref{delta}) at  their best fit values,  we have numerically  calculated the CPT asymmetry for various scenarios of $L$ and $E$, as presented in   Table (\ref{tabcpt}).   The values of $A_{\mu e}^{CPT}$  calculated using the approximate expressions of the probabilities have also been presented in the table.

\begin{table}[h]
\caption{CPT asymmetry $A_{\mu e}^{CPT}$  for various scenarios of $L$ and  $E$. All  other input parameters are kept at their best fit values given in the equations (\ref{masses}) - (\ref{delta}).}
\begin{tabular}{|cllll|} \hline
 Scenario &   L (Km) &  E (GeV)    
 & $A_{\mu e}^{CPT}$  & $A_{\mu e}^{CPT}$  \\ 
& & & exact & approximate \\  \hline
S1 & 300    & 1.0 & 0.058 & 0.058  \\
S2 & 1300 & 3.5    & 0.31 & 0.30\\
S3 &  2300 &  5.0 &  0.63 & 0.62\\
S4 & 3000  &  7.0   &  0.73 & 0.72\\  
\hline
\end{tabular}

\label{tabcpt}

\end{table}

We observe from the Table (\ref{tabcpt}), that the magnitude of the CPT asymmetry in these experiments is not small, particularly for baselines greater than 1000\,Km, the asymmetry is large enough.  However, it should be borne in mind that due to the oscillatory behavior of the CPT asymmetry, the magnitude of the asymmetry may vary greatly on slightest variation of the neutrino energy $E$ and/or the baseline length $L$. Therefore it is more appropriate to graphically show the variation of   $A_{\mu e}^{CPT}$ with neutrino energy $E$.

  In Figure (\ref{fige}), we have plotted the approximate   as well as the exact   magnitude of 
 $A_{\mu e}^{CPT}$ as  function of neutrino energy  $E$ for the four baselines given in Table (\ref{tabcpt}). The upper limit of  energy range chosen corresponds to the range available to LENF. All other input parameters have been kept at their best fit values given in equations (\ref{masses}) - (\ref{delta}). It may be mentioned that the neutrinos have been assumed to follow  normal hierarchy of masses throughout this work.
We observe that the peak value of the CPT asymmetry $A_{\mu e}^{CPT}$ increases with increasing neutrino energy. This behavior is expected, as the matter effects increase with the neutrino energy. Further, for a given energy,  $A_{\mu e}^{CPT}$ is maximum for S4 and minimum for S1, implying that  $A_{\mu e}^{CPT}$ increases with baseline length.  On comparing the four plots in Figure(\ref{fige}), we find that  the rise in $A_{\mu e}^{CPT}$ per unit increase in energy is maximum for S4, implying that longer the baseline, more is the sensitivity of  $A_{\mu e}^{CPT}$ towards the neutrino energy.
Thus, it may be inferred that extrinsic CPT violation may have a  significant magnitude for long baseline neutrino oscillation experiments.

\begin{figure} 
\caption{(Color online). CPT asymmetry $A_{\mu e}^{CPT}$  as function of neutrino energy $E$ and baseline length $L$. The dots correspond to  $L$ and  average energy $E$ of the experimental scenarios given in  Table (\ref{tabcpt}),  while an assumed energy spread of 20$\%$ in the beam is indicated by the error bars. All other input parameters are kept at their best fit values given in equations (\ref{masses}) - (\ref{delta}).}
\includegraphics [width=.5\textwidth] {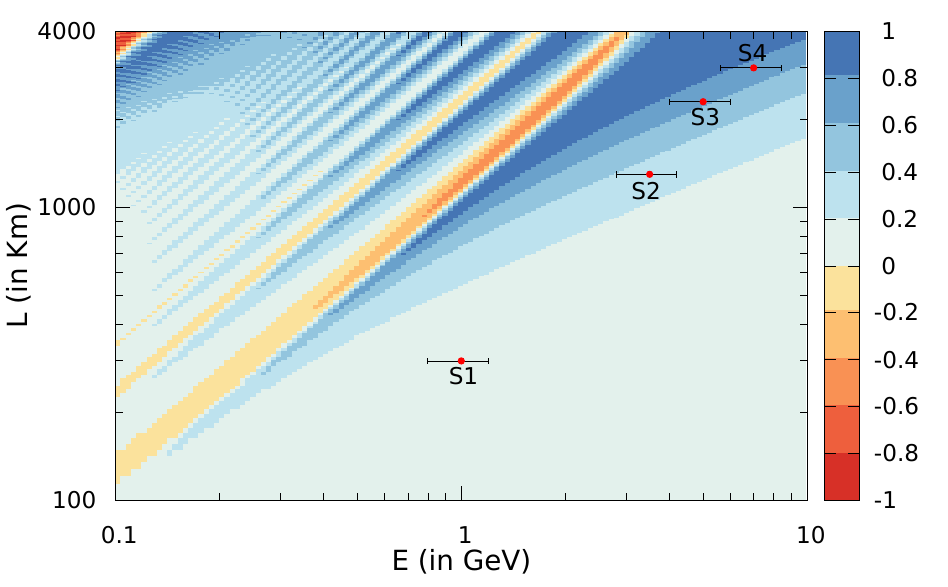}
\label{le}
\end{figure}

As far as the validity of the approximate analytical expressions is concerned, the plots reveal that the agreement between the approximate and the exact calculations is better at higher energies in comparison to lower energies. This is due to the reason that the approximate expressions for the probabilities given in equations (\ref{pmue}) and (\ref{pemubar}) are valid only when $L/E \ll 10^4$ Km/GeV, i.e. far from the  $L/E$ region where the low frequency solar oscillations become dominant. Therefore, one must be careful about the region, where the approximate analytic formulas may be applied.

In figure (\ref{le}), we present the exact calculations of  $A_{\mu e}^{CPT}$ as functions of $E$ and $L$.
The dots indicate the baseline length $L$ and the average neutrino  energy $E$ for various experimental scenarios given in  Table (\ref{tabcpt}).  An  assumed energy spread of 20$\%$  in the beam is indicated by the error bars.          
   It is clear from the figure that  
$A_{\mu e}^{CPT}$ is maximum  at upper right corner, where both $E$ and $L$ are large. 
 At the lower right corner $A_{\mu e}^{CPT}$ is too small to be of significance. 
 The effect of extrinsic CPT violation is maximum for S3 and S4, where it is between  0.6 - 0.8\,.   For S1 and S2 it is less than  0.4\,.
Further, it may be seen that $A_{\mu e}^{CPT}$ values will not change significantly within the whole spread of energy for S1 and S2, however for S3 and S4, $A_{\mu e}^{CPT}$ may become  larger at the lower end of the energy dispersion.
Thus, for these experiments the effect of extrinsic CPT violation is not only large, but will further increase at  neutrino energies which are lower than the average value. However, these results should be interpreted rather carefully, since in real experiments, the detectors have a finite energy resolution, very fast oscillations at low energies can not be resolved. Therefore, one should consider probabilities averaged over the energy resolutions of the detectors.
Moreover, to make any final comment about the magnitude of the CPT asymmetry in any experiment, it is of utmost importance to mention that the two CPT conjugate channels should be compared in terms of neutrino event rates, which apart from the oscillation probabilities, also depend on neutrino-nucleon cross section and initial flux of neutrinos. In the present work, however, we confine our analysis to the study of   oscillation probabilities only. The analysis with event rates will be discussed in a future publication.

\begin{figure} [b]
\caption{(Color online). CPT asymmetry $A_{\mu e}^{CPT}$  as function of  CP violation phase $\delta_{CP}$ for the scenarios S1, S2, S3 and S4. The thick black curve corresponds to $L=3000$\,Km and $E=0.5$\,GeV.}
 \includegraphics[width=.5\textwidth] {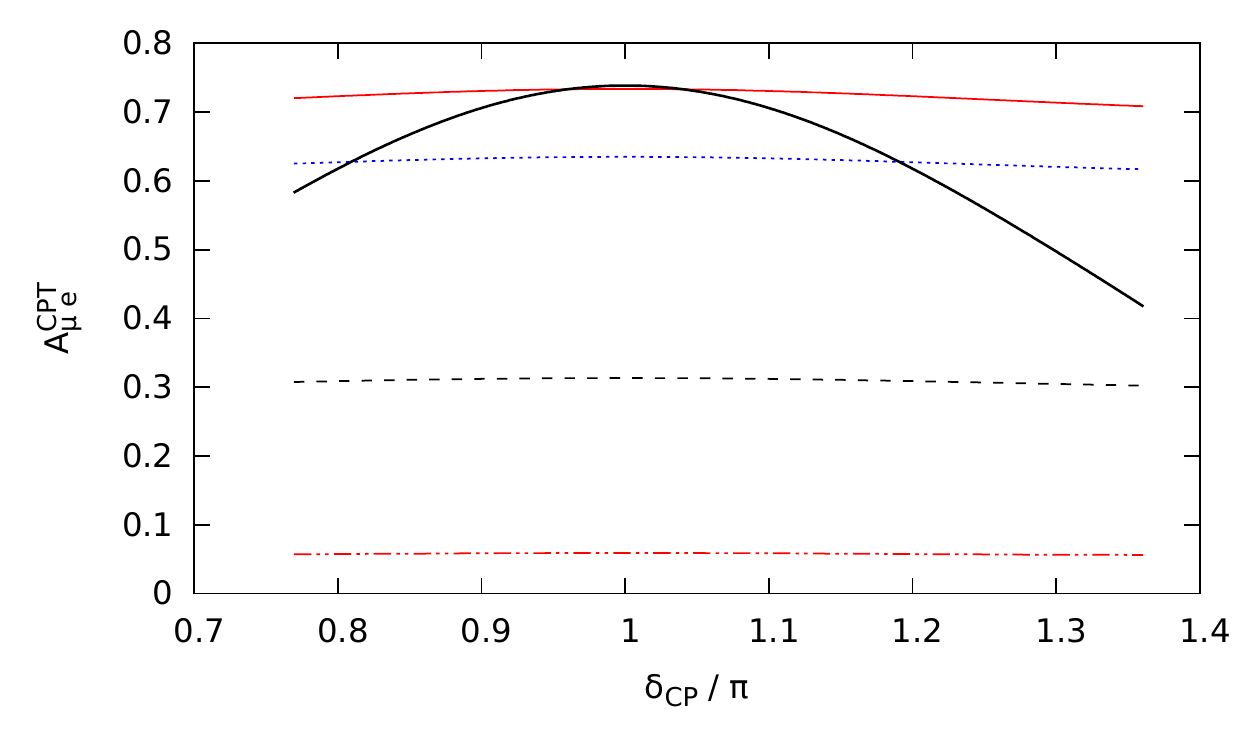}
 \label{del}
\end{figure}

It is interesting to note that $A_{\mu e}^{CPT}$ is very sensitive to    variations in $\theta_{23},~ \theta_{13}$ and 
$\Delta m_{23}^{2}$, while   variations in $ \theta_{12}$, $\Delta m_{12}^{2}$ and $\delta_{CP}$ hardly affect $A_{\mu e}^{CPT}$. Our analysis shows that the sensitivity of $A_{\mu e}^{CPT}$ towards $\theta_{23},~ \theta_{13}$ and 
$\Delta m_{23}^{2}$ increases with increasing baseline length and decreases with increasing values of the  average neutrino energy. However, at longer baseline lengths the effect of $L$ is more pronounced than the effect of energy.
 Therefore, despite having a high value of average energy $E$, S4 set up has highest sensitivity towards variation in $\theta_{23},~ \theta_{13}$ and 
$\Delta m_{23}^{2}$ followed by S3, S2 and S1, in that order. For example, for S4, at the upper limit of  $\theta_{13}$,  $A_{\mu e}^{CPT}$ increases from 0.3 to 0.6, all other parameters being at their mean values.
These results assume significance in the wake of the fact that the precision in the determination of ${\rm sin}^2\,\theta_{13}$ is less in comparison to other parameters.
Further, it is worth noting that though   $A_{\mu e}^{CPT}$ values change very little with  $\delta_{CP}$ in the $L/E$ region relevant for various experimental scenarios discussed in the text, in the low $E$ and longer $L$ region,  $A_{\mu e}^{CPT}$ varies significantly with  $\delta_{CP}$ as shown by the thick black curve in figure (\ref{del}), which  corresponds to $L=3000$\,Km and $E=0.5$\,GeV. 
All other lines in the figure, corresponding to the experimental scenarios given in Table  (\ref{tabcpt}), are almost insensitive to variations in $\delta_{CP}$. This is due to the reason that the probability itself is large in the low $E$ region, and hence also is more sensitive to the variation of  $\delta_{CP}$.
Thus it may be said that the magnitude of CPT asymmetry is sensitive to the magnitude of CP violation in the high $L/E$ region.

\section{Conclusions}
In conclusion, we have investigated  the implications of matter induced CPT violation effects on  the transition probabilities for neutrino oscillations in some  scenarios of  long baseline electron neutrino appearance experiments, in a low energy neutrino factory like setup. We find that the magnitude of CPT asymmetry $A_{\mu e}^{CPT}$ in these experiments is not ignorable, particularly for baselines greater than 1000\,Km, the asymmetry is large enough. The
 peak value of the CPT asymmetry increases with increasing neutrino energy as well as with baseline length. We have also examined the dependence of CPT violating    asymmetry  on the oscillation parameters like mixing angles, mass squared differences as well as on the Dirac CP violating phase for these long baseline experiments. We observe that  
$A_{\mu e}^{CPT}$ is very sensitive to    variation in $\theta_{23},~ \theta_{13}$ and 
$\Delta m_{23}^{2}$, while the variations in $ \theta_{12}$, $\Delta m_{12}^{2}$  hardly affect $A_{\mu e}^{CPT}$.  Although,  $A_{\mu e}^{CPT}$ values change very little with  $\delta_{CP}$,  we observe that in the low $E$ and longer $L$ region, $A_{\mu e}^{CPT}$ varies significantly with  $\delta_{CP}$, suggesting that the magnitude of CPT asymmetry is sensitive to the magnitude of CP violation in the high $L/E$ region. It is  suggested that the experimental collaborations should investigate the effects of extrinsic CPT violation in their respective experimental setups.

 \begin{acknowledgments}
 M.S. would like to thank the Chairman, Department of Physics for providing facilities to work.
M.R. is supported by the UGC, Govt. of India, under the Research Award Scheme (No.F.30-39/2011(SA-II)).
\end{acknowledgments}

\pagebreak

\end{document}